\documentclass{elsart1p}
\usepackage{amssymb,amsmath}
\usepackage{graphicx}
\newcommand{\bra}{\langle}
\newcommand{\ket}{\rangle}
\if@defined\half
\@else
\newcommand{\half}{\frac{1}{2}}
\fi

\newcommand{\One}{1\kern-4.5pt1}

\if@defined\implies
\renewcommand{\implies}{\Longrightarrow}
\@else
\newcommand{\implies}{\Longrightarrow}
\fi

\renewcommand{\vec}[1]{\overrightarrow{#1}}

\newcommand{\gam}{\gamma}

\newcommand{\psibar}{\overline{\psi}}

\newcommand{\qctd}{QC$_2$D}
\newcommand{\ctcqcd}{\cite{Hands:2006ve}}
\newcommand{\cdse}{\cite{Roberts:2000aa,Nickel:2006vf}}
\newcommand{\cgluon}{\cite{Leinweber:1998im,Cucchieri:2007md,Bogolubsky:2007ud}}

\newcommand{\ckstvz}{\cite{Kogut:2000ek}}

\begin{document}
%\bibliographystyle{h-elsevier3}

% declarations for front matter
\begin{frontmatter}
\title{Quarks and gluons in dense two-colour QCD}

\author{Jon-Ivar Skullerud}
\address[NUIM]{Department of Mathematical
  Physics, NUI Maynooth, County Kildare, Ireland}
%  \thanks{Speaker}, S.~J.~Hands\address[Swan]{Department of Physics,
%  Swansea University, Singleton Park, Swansea SA2 8PP, Wales, UK}

\begin{abstract}
We compute quark and gluon propagators in 2-colour QCD at large baryon
chemical potential $\mu$.  The gluon propagator is found to be
antiscreened at intermediate $\mu$ and screened at large $\mu$.  The
quark propagator is drastically modified in the superfluid region as a
result of the formation of a superfluid gap.
\end{abstract}

%\begin{keyword}

%\PACS
%\end{keyword}
\end{frontmatter}

\section{Introduction}

Determining the phase diagram of QCD at large baryon density and small
temperatures remains one of the outstanding problems of strong
interaction physics. %  This problem is of both theoretical and
%phenomenological interest: on the theoretical side, an exceptionally
%rich phase structure may be present, while the phenomenological
%interest is spurred by the possibility that some of these phases may
%be present in compact stars, and may have observable consequences.
%
Direct lattice simulations of QCD at high density and low temperature are
hindered by the sign problem, so alternative approaches are
required. One such approach is to study QCD-like theories which may
be simulated on the lattice, and apply the lessons learnt from these
theories to the case of real QCD.  Foremost among these theories is
QCD with gauge group SU(2) (\qctd).
Medium modifications of quark and gluon propagators is one topic where
\qctd\ may directly inform real QCD calculations.  The gluon
propagator is used as input into the gap equation for the superfluid
gap at high density, and nontrivial medium modifications may
significantly alter the results.  %The quark propagator encodes
%information about effective quark masses and gap parameters, while
First-principles results for gluon and quark propagators together can
be used to check the assumptions going into dense QCD calculations in
the Dyson--Schwinger equation framework \cdse.

%\section{FORMULATION}

We will be using $N_f=2$ degenerate flavours of Wilson fermion, with a
diquark source $j$ included to lift low-lying eigenvalues and study
diquark condensation without uncontrolled approximations.  The fermion
action can be written
\begin{equation}
S_F = \begin{pmatrix}\psibar_1 & \psi_2^T\end{pmatrix}
 \begin{pmatrix} M(\mu) & j\gam_5 \\ -j\gam_5 & M(-\mu)\end{pmatrix}
 \begin{pmatrix}\psi_1 \\ \psibar_2^T\end{pmatrix}
 \equiv \overline\Psi\mathcal{M}(\mu)\Psi\,,
\end{equation}
where $M(\mu)$ is the usual Wilson fermion matrix with chemical
potential $\mu$.  It satisfies the symmetries
%
%\begin{align}
$KM(\mu)K^{-1} = M^*(\mu)\,,
\gam_5M^\dagger(\mu)\gam_5 = M(-\mu)\,,$
%\end{align}
%
with $K=C\gam_5\tau_2$.  The first of these is the Pauli--G\"ursey
symmetry. The inverse of $\mathcal{M}$ is the Gor'kov
propagator,
\begin{equation}
\mathcal{G}(x,y) = \mathcal{M}^{-1} =
 \begin{pmatrix}
 \bra\psi_1(x)\psibar_1(y)\ket & \bra\psi_1(x)\psi_1^T(y)\ket\\
 \bra\psibar_2^T(x)\psibar_1(y)\ket & \bra\psibar_2^T(x)\psi_1^T(y)\ket
 \end{pmatrix}
 =
 \begin{pmatrix}S(x,y) & T(x,y) \\ \bar{T}(x,y) & \bar{S}(x,y)
 \end{pmatrix} \,.
\end{equation}
The components $S$ and $T$ denote normal and anomalous propagation
respectively.%  The Gor'kov propagator has the symmetry properties
%%
%\begin{gather}
% K\mathcal{G}K^{-1} =
% \begin{pmatrix} S^* & -T^* \\ -\bar{T}^* & \bar{S}^*\end{pmatrix}
%   \,,\\
%\bar{S}(x,y) = -S(y,x)^T\,,\qquad T(x,y)=T(y,x)^T\,,\qquad
% \bar{T}(x,y) = \bar{T}(y,x)^T\,.
%\end{gather}
%%
%We will also write the inverse propagator as
%%
%\begin{equation}
%\mathcal{G}^{-1} = \begin{pmatrix} N & A \\ \bar{A} & \bar{N}
%\end{pmatrix}\,,
%\end{equation}
%%
%which has the same symmetry properties as $\mathcal{G}$.
The normal (diagonal) part $N$ of the inverse Gor'kov propagator can in general be
written in terms of four momentum-space form factors,
\begin{equation}
N(p) = \not\vec{p}A(\vec{p}^2,p_4) + B(\vec{p}^2,p_4)
 + \gam_4(p_4-i\mu)C(\vec{p},p_4) +
 i\gam_4\not\vec{p}D(\vec{p}^2,p_4)\,.
 \label{formfactors}
\end{equation}
Analogous form factors $S_a, S_b, S_c, S_d$ can be written for the
propagator $S$.  In \qctd\ the Pauli--G\"ursey symmetry ensures that
all form factors are purely real.  %The structure of the anomalous
%propagator depends on the pattern of diquark condensation.  
Assuming
that the condensation occurs in the colour singlet channel with
quarks of unequal flavour, the anomalous propagator can be written
as $T(p) = T'(p)C\Gamma\tau_2$ (and similarly for the anomalous part
$A(p)$ of the inverse propagator), where
$\Gamma=\gam_5$ for condensation in the scalar ($0^+$) channel. %and
%pseudoscalar ($0^-$) channel respectively.  %Spin-1 condensation
%leads to more complicated structures, but is energetically
%disfavoured compared to spin-0 condensation and will not be
%considered here.  
The remaining spin structure can be written in
terms of form factors $T_a, T_b, T_c, T_d$ analogous to
\eqref{formfactors}.  The form factors $\phi_a, \phi_b, \phi_c,
\phi_d$ for $A'(p)$ are the gap functions.

The gluon propagator in presence of a chemical potential in Landau
gauge may be decomposed into an magnetic and electric form factor,
\begin{equation}
D_{\mu\nu}(\vec{q},q_0) = P^T_{\mu\nu}D_M(\vec{q}^2,q_4^2) +
P^E_{\mu\nu}D_E(\vec{q}^2,q_4^2) \,.%+ \xi\frac{q_\mu q_\nu}{q^4}\,.
\end{equation}
The projectors $P^T_{\mu\nu}(q), P^E_{\mu\nu}(q)$ are both
4-dimensionally transverse, and are spatially transverse and
longitudinal respectively.
%We will be working in Landau gauge ($\xi=0$), so the 4-d longitudinal
%piece vanishes.

\section{Results}

We have generated gauge configurations on two lattices: a ``coarse''
lattice with $\beta=1.7, \kappa=0.178, V=8^3\times16$, and a 
``fine'' lattice with $\beta=1.9, \kappa=0.168, V=12^3\times24$.  The
lattice spacings are 0.23fm and 0.18fm respectively, while
$m_\pi/m_\rho=0.8$ in both cases.
A range of chemical
potentials $\mu$ were used with diquark source $aj=0.04$, while additional
configurations were generated with $aj=0.02, 0.06$ for selected values
of $\mu$.
Results for the gluon propagator on the coarse lattice have been
presented in \ctcqcd; we will supplement these here with results from
the fine lattice.  On both lattices, an onset transition to a phase
with nonzero baryon density and diquark condensate was found at
$\mu_o\approx m_\pi/2$, while BCS-like scaling of energy density, baryon
density and diquark condensate was found at higher $\mu$.  On the
coarse lattice the crossover to BCS-like scaling was associated with a
nonvanishing Polyakov loop $L$, indicating a coincident deconfinement
transition \ctcqcd.  On the fine lattice, this transition has been
pushed to considerably larger $\mu$ \cite{Hands:priv}.

\begin{figure}[t]
\begin{center}
\includegraphics*[width=0.45\textwidth]{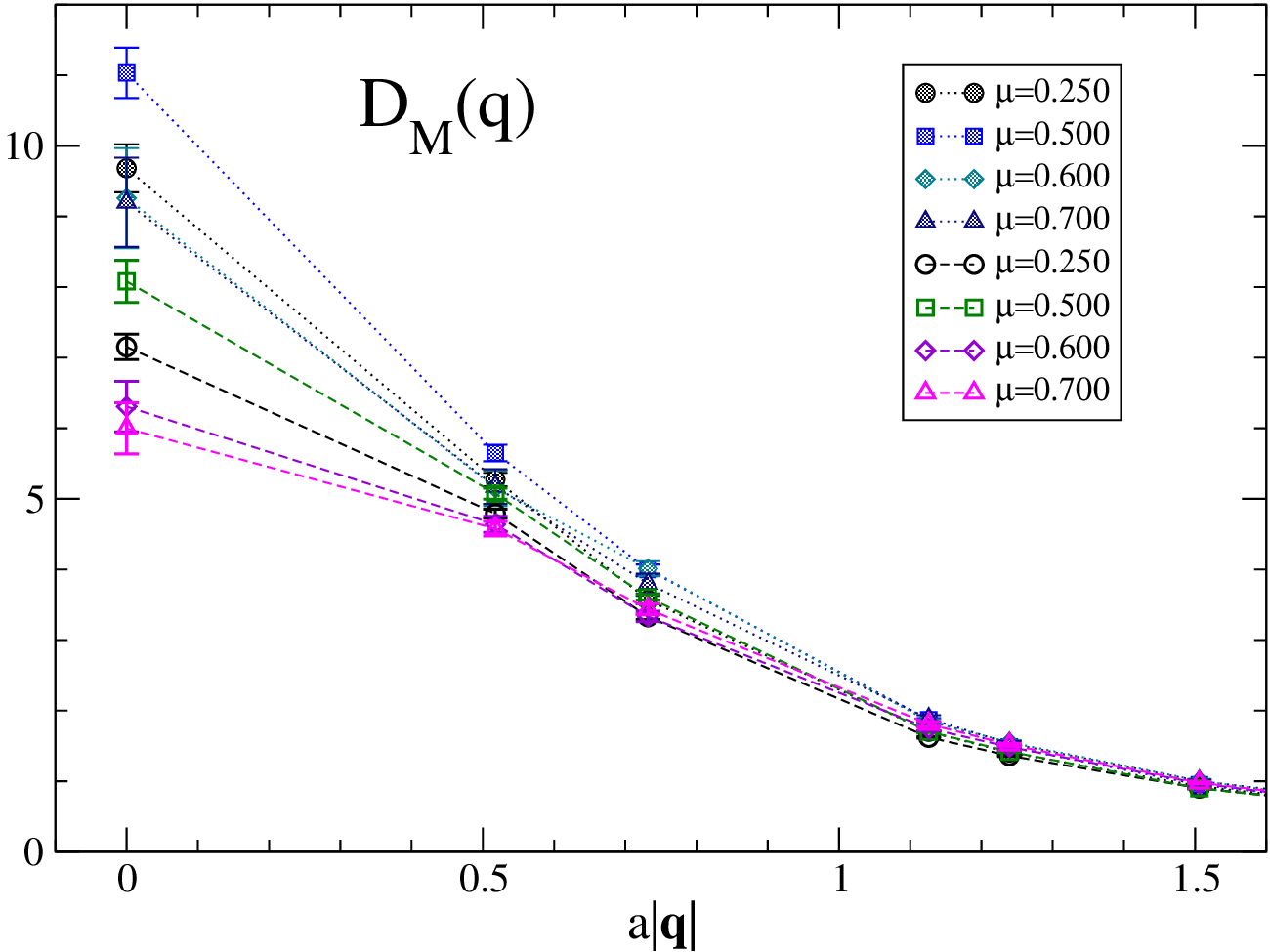}
\includegraphics*[width=0.45\textwidth]{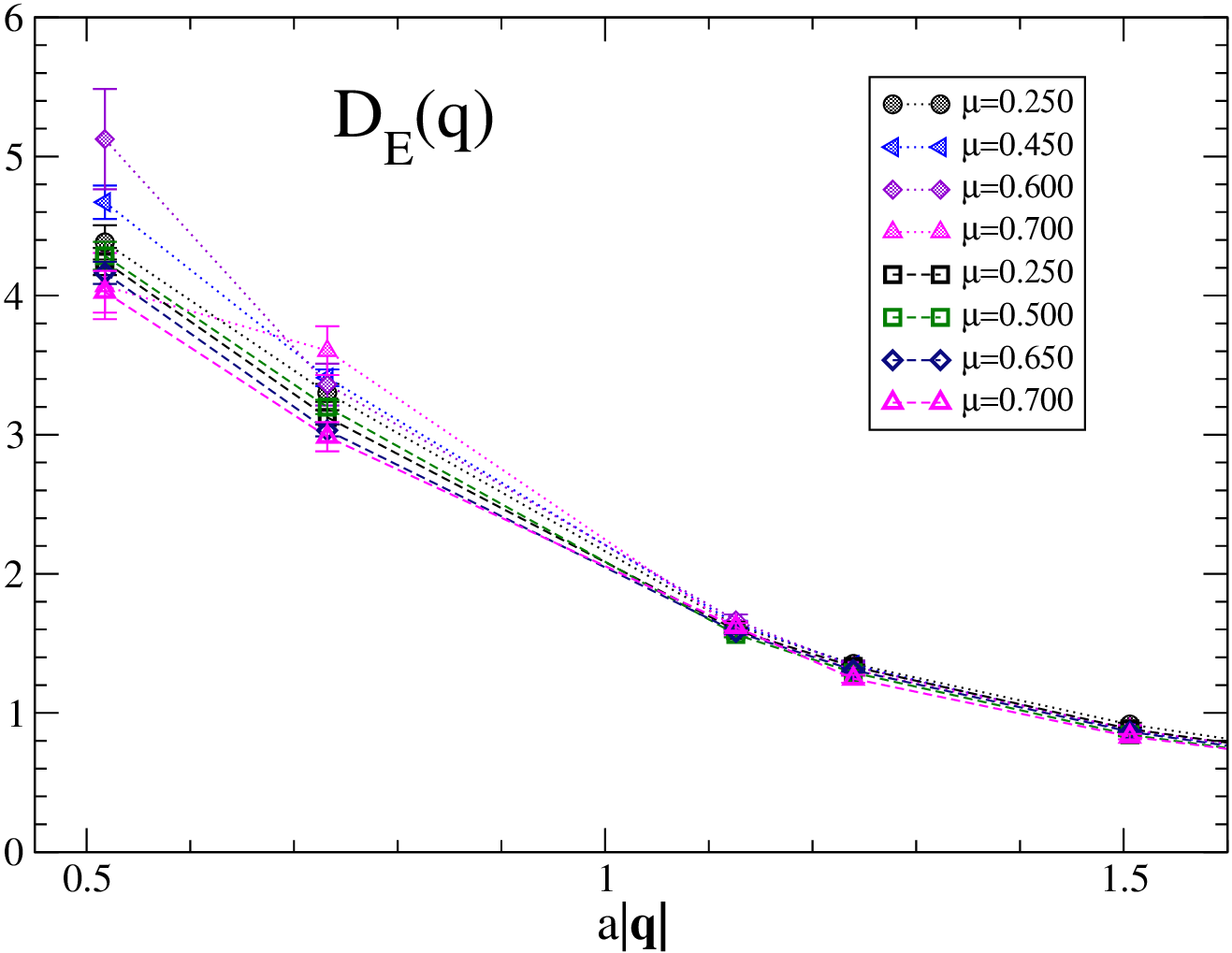}
\caption{Magnetic (left) and electric (right) gluon propagator on the
  fine lattice.  The filled symbols
  represent the lowest Matsubara mode ($q_4=0$), while the open symbols
  represent the first nonzero Matsubara mode.}
\label{fig:gluon}
\end{center}
\end{figure}
Figure~\ref{fig:gluon} shows the gluon propagator as a function of
spatial momentum $|\vec{q}|$ for the two lowest Matsubara frequencies,
on the fine lattice.  In all cases, the propagator at the lowest chemical
potential $\mu$ shown is consistent with the vacuum propagator.  On
the coarse lattice \ctcqcd\ both magnetic and electric propagator are strongly
screened at large $\mu$, while they are enhanced at low momentum in
the intermediate-density region.  %The static ($q_0=0$) magnetic gluon
%propagator turns out to have a surprisingly strong dependence on the
%diquark source, which counteracts the infrared suppression at large
%$\mu$ as $j\to0$, but does not remove it completely.  
On the fine
lattice, the non-static modes of the magnetic propagator remain
screened, while the static mode and the electric propagator experience
much weaker modifications.  The enhancement in the intermediate,
superfluid region remains.  The weaker screening may be linked to the
deconfinement transition occuring at much higher $\mu$, implying that
the gluon is not screened by coloured quark degrees of freedom at
these densities.

\begin{figure}[hbt]
\begin{center}
\includegraphics*[width=0.45\textwidth]{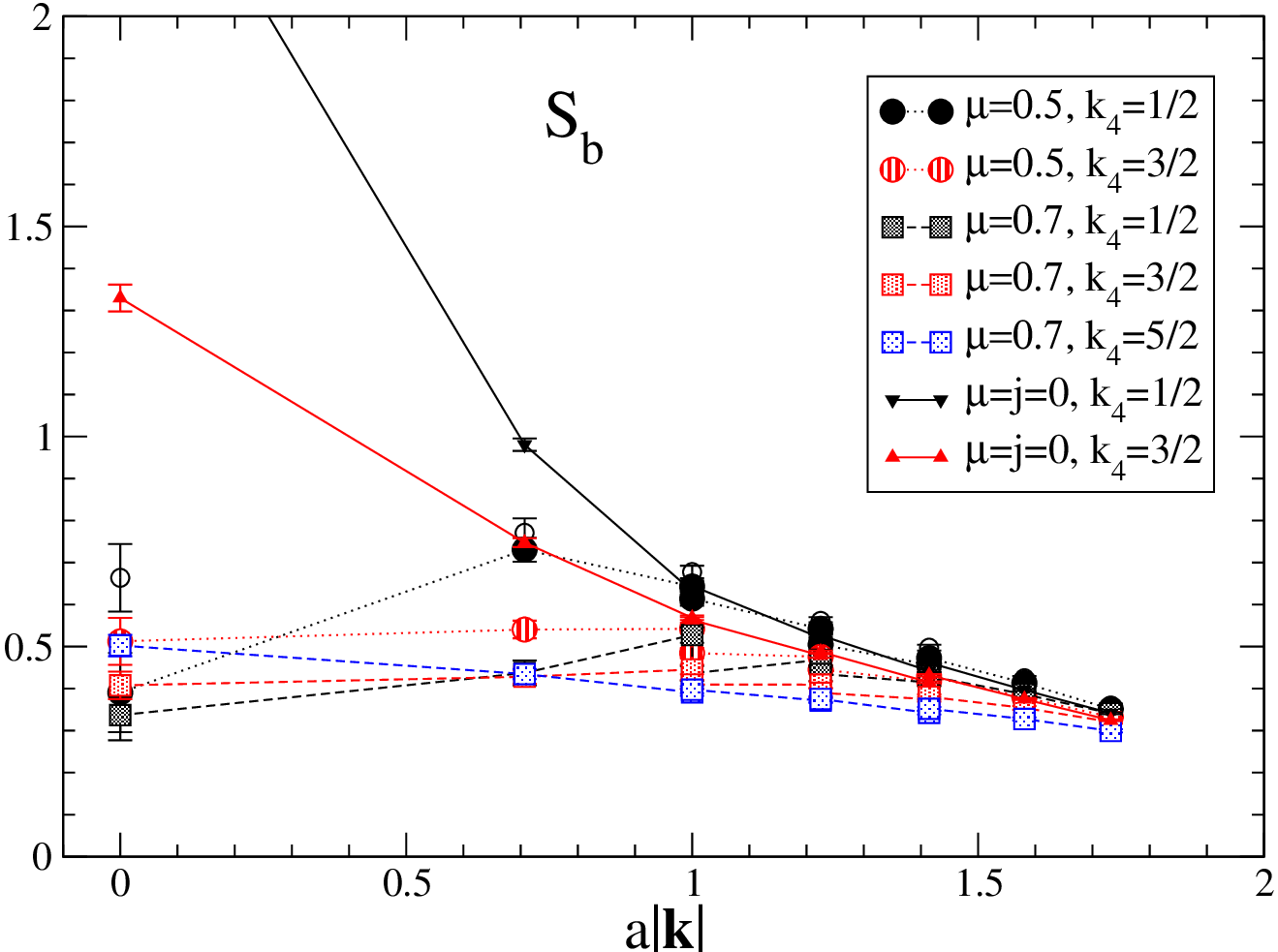}
\includegraphics*[width=0.45\textwidth]{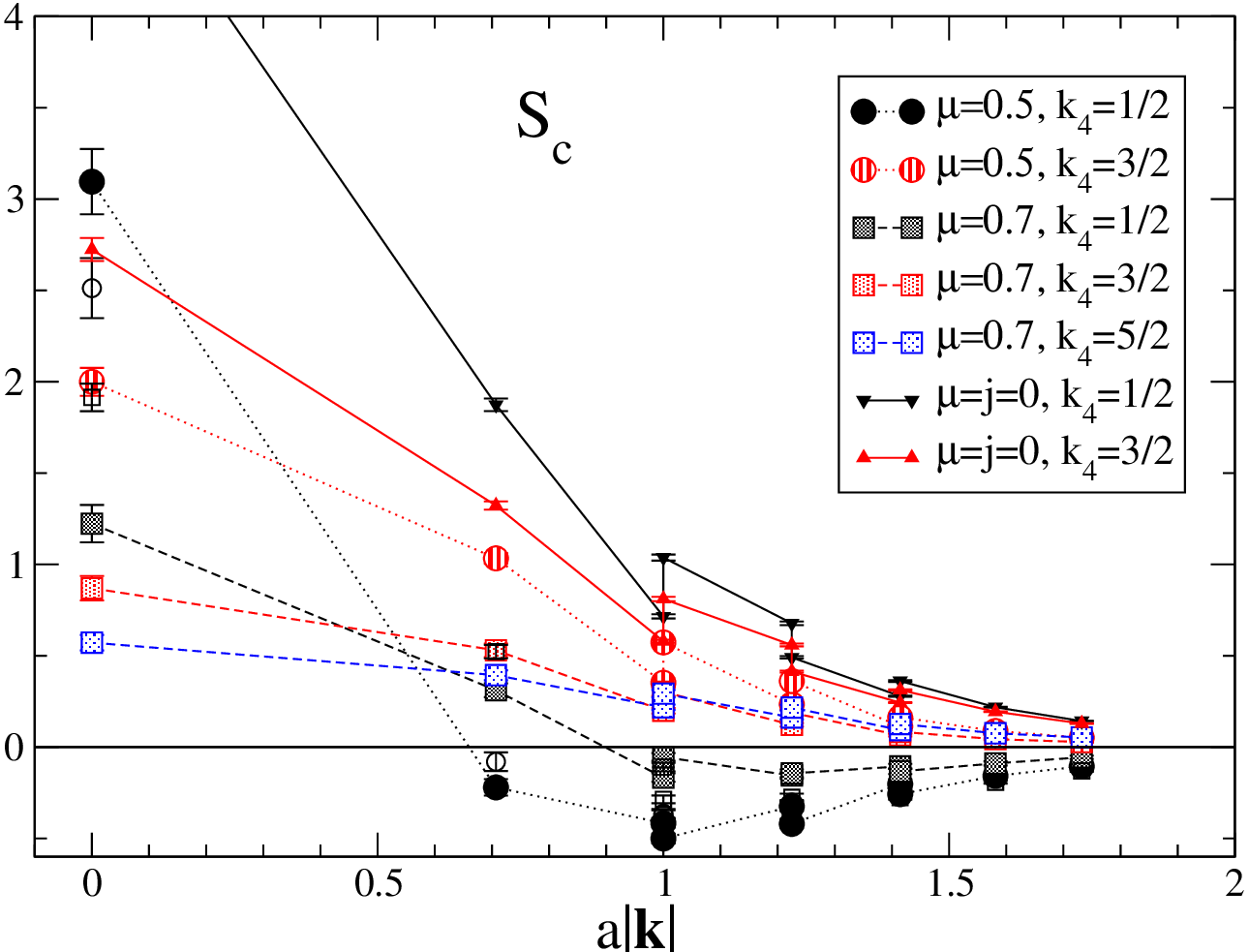}
\end{center}
\caption{The scalar (left) and temporal-vector (right) part of the normal
  quark propagator, on the coarse lattice.  The filled symbols are for
  $aj=0.02$, while the open symbols are $aj=0.04$.  The vacuum
  ($\mu=j=0$) propagator is also shown for comparison.}
\label{fig:normal}
\end{figure}
Figure~\ref{fig:normal} shows the scalar part $S_b$ and
temporal-vector part $S_c$ of
the normal quark propagator for $a\mu=0.5, 0.7$ on the coarse
lattice.  These both exhibit dramatic medium modifications.  The
scalar propagator $S_b$ is strongly suppressed in the superfluid
phase, suggesting a drastic reduction in the in-medium effective quark
mass.  This is linked to the appearence of the diquark condensate: the
chiral condensate rotates into the diquark condensate in the
superfluid phase \ckstvz.  We would therefore expect to find the
missing strength in the anomalous propagator.
The lowest Matsubara mode of  $S_c$
becomes negative at intermediate momenta, approaching zero from below
at high momenta.  This is a signal of a gap, and the location of the
zero crossing in the $k_0\to0$ limit can be used to determine the
Fermi momentum.  In accordance with this, the zero crossing moves to
larger $|\vec{k}|$ as $\mu$ increases.  The spatial-vector propagator
$S_a$ experiences smaller modifications, while the tensor part $S_d$
is found to be consistent with zero.  Preliminary results from the
fine lattice confirm this picture.

\begin{figure}[hbt]
\includegraphics*[width=0.45\textwidth]{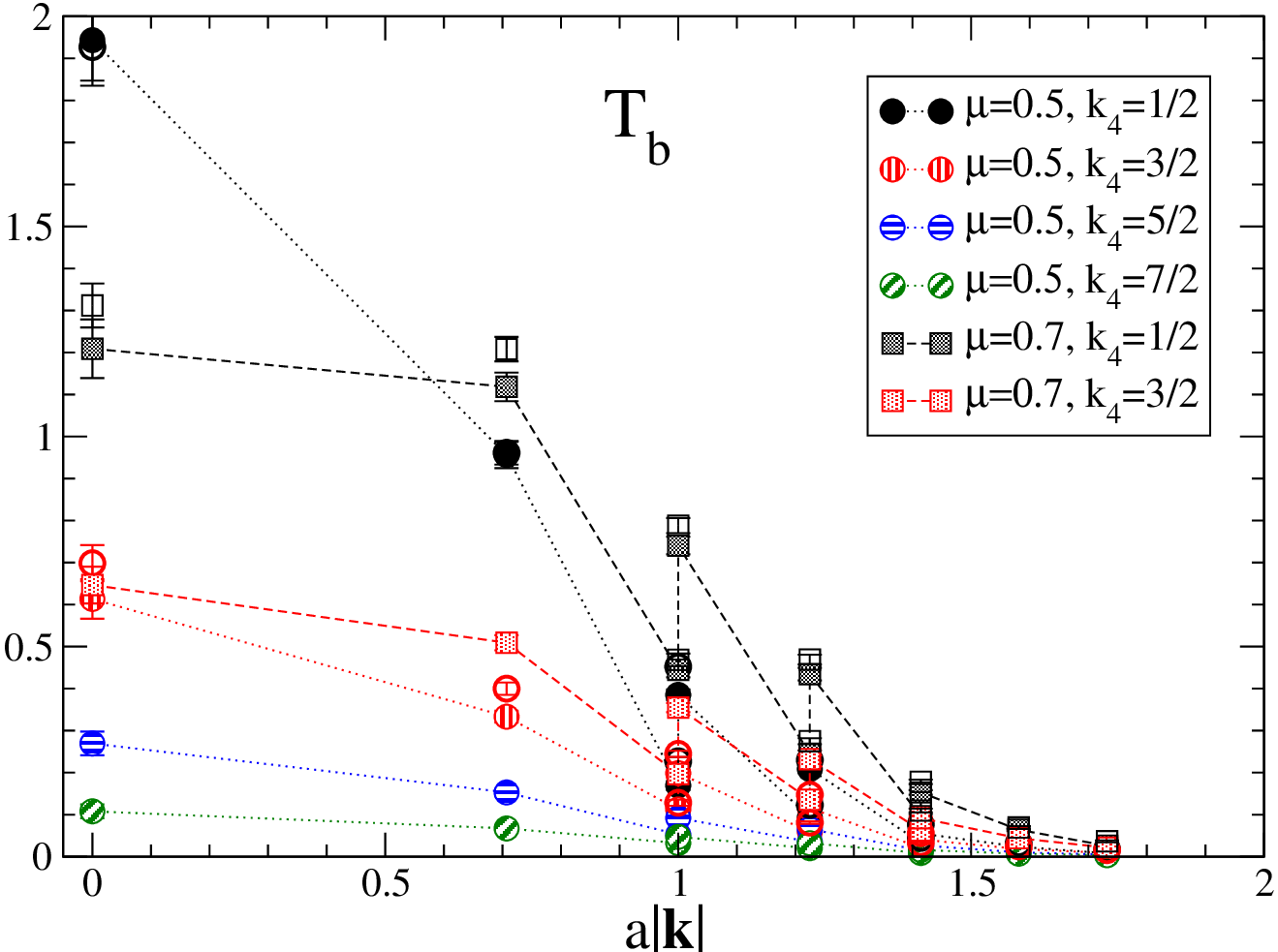}
\includegraphics*[width=0.45\textwidth]{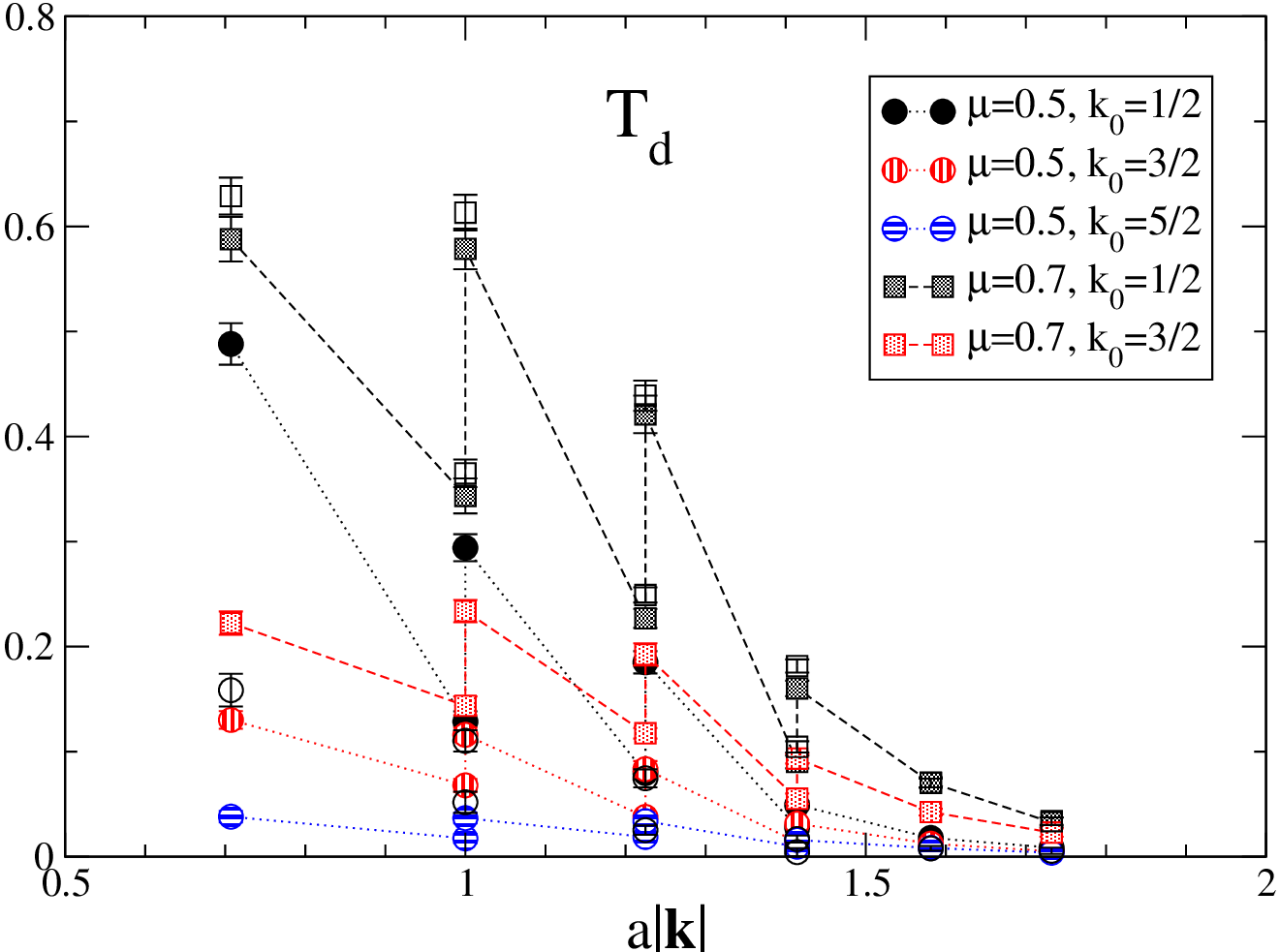}
\caption{The scalar (left) and tensor (right) part of the anomalous
  quark propagator, on the coarse lattice.  The filled symbols are for
  $aj=0.02$, while the open symbols are $aj=0.04$.}
\label{fig:anomalous}
\end{figure}
Figure~\ref{fig:anomalous} shows the nonzero components of the
anomalous Gor'kov propagator.  The dominant part is, as expected, the
scalar part $T_b$, but a clear signal is also found for the tensor
part $T_d$.  There is also a non-zero signal for the lowest Matsubara
frequency of the temporal-vector propagator $T_c$.  %We see that
Neither the normal nor the anomalous propagator depend strongly on the
diquark source term $j$.  %, indicating that we can safely take the
%$j\to0$ limit.  
Lattice artefacts are however quite large, as
indicated by the discrepancy between the two points at $a|\vec{k}|=1$
%(and also between the points at $a|\vec{k}|=\sqrt{2}$).  
Quantitative
results will require analysis of the quark propagator on the fine
lattice.

\section{Discussion and outlook}

We have found substantial modifications of both gluon and quark
propagators in the dense medium.  Both electric and magnetic gluon
propagator are screened at large $\mu$, but a more quantitative
analysis, including extrapolation to $j=0$ and fits to determine
screening masses, will be necessary to draw firm conclusions.  Note
that the gluon propagator in known to be infrared suppressed in the
vacuum \cgluon, so even without further medium modifications the
static magnetic gluon propagator will be screened, in contradiction to
any perturbative approximation.  %In order to reproduce the correct
%quark--quark interaction, it may also be important to incorporate
%nonperturbative structure in the quark--gluon vertex \cqqg.

The dramatic modifications seen in the quark propagator are directly
related to the appearance of a diquark gap.  Further quantitative
studies of this will include determining the Fermi momentum $p_F$ from
$S_c(p_F,k_4=0)=0$ and determining the size of Cooper pairs from the
anomalous propagator, to study the BEC--BCS crossover in more detail.

\section*{Acknowledgments}

I wish to thank Simon Hands for his collaboration in this research,
and Dominik Nickel for very fruitful discussions.

%\bibliography{gluon,density,qcd}

\end{document}